\documentclass[preprint, 12pt]{elsarticle}
\usepackage{graphicx} 
\usepackage{amsmath,amsfonts,amssymb,amsthm}
\usepackage{mathtools}
\usepackage{color}
\usepackage{booktabs}
\usepackage[T1]{fontenc}
\usepackage{lineno}
\usepackage[utf8]{inputenc}

\begin{document}

\begin{frontmatter}

\title{Emergence of Chimeras States in One-dimensional Ising Model with Long-Range Diffusion}
\author[1]{Alejandro de Haro García}
\ead{alexdeharo269@gmail.com}
\author[1]{Joaquín J. Torres\corref{cor1}}
\ead{jtorres@onsager.ugr.es}
\affiliation[1]{Institute ``Carlos I'' for Theoretical and Computational Physics, and Department of Electromagnetism and Physics of the Matter,
  University of Granada, E-18071 Granada, Spain}
\cortext[cor1]{Corresponding author}

\begin{abstract}
In this work, we examine the conditions for the emergence of chimera-like states in Ising systems. We study an Ising chain with periodic boundaries in contact with a thermal bath at temperature $T, $ that induces stochastic changes in spin variables.  To capture the non-locality needed for chimera formation, we introduce a model setup with non-local diffusion of spin values through the whole system. More precisely, diffusion is modeled through spin-exchange interactions between units up to a distance $R$, using Kawasaki dynamics. This setup mimics, e.g., neural media, as the brain, in the presence of electrical (diffusive) interactions. We explored the influence of such non-local dynamics on the emergence of  complex spatiotemporal synchronization patterns of activity. Depending on system parameters we report here for the first time chimera-like states in the Ising model,  characterized by relatively stable moving domains of spins with different local magnetization. We analyzed the system at $T=0,$ both analytically and via simulations and computed the system’s phase diagram, revealing rich behavior: regions with only chimeras, coexistence of chimeras and stable domains, and metastable chimeras that decay into uniform stable domains. This study offers fundamental insights into how coherent and incoherent synchronization patterns can arise in complex networked systems as it is, e.g., the brain.

\end{abstract}

\end{frontmatter}


\section{Introduction}
Over the last decades, significant advances have been made in the study of complex systems and on the intriguing emergent phenomena associated with them \cite{sethna2006,mitchell2009,complex2009, munoz2018, complex2020}. Complex systems consist, in general, of many interconnected components whose interactions give rise to collective behaviors that are often nonlinear, emergent, and difficult to predict, although similar phenomenology can also appear in low-dimensional systems presenting, for instance, chaotic behavior \cite{CRUTCHFIELD1994,chialvo2010,strogatz2018}. Complex systems are found across various domains, including biology (e.g., neural networks, ecosystems), physics (e.g., fluid turbulence, spin systems), and society (e.g., economies, social dynamics). The study of complex systems aims to understand how local interactions can lead to global patterns and behaviors, often through the lens of network theory, nonlinear dynamics, and statistical mechanics \cite{mitchell2009}. 

Within this broad field, the so-called {\em chimera} states represent a striking example of emergent complex behavior. First described in systems of identically coupled oscillators by Kuramoto and Battogtokh \cite{kuramoto2002},  chimera states  are characterized by regions of synchronized  (coherent) and desynchronized (incoherent) activity coexisting in time -- a paradoxical outcome given the system's uniform structure -- and are associated to the non-local features of the elements coupling and interactions. This phenomenon was later named and formalized by Abrams and Strogatz \cite{abramsstrogatz2004}, who theoretically demonstrated its relevance and the conditions for its appearance and stability,  which moreover has been related with factors like the system's size, the nature of the coupling, and the specific network structure of the system \cite{wolfrum2011,Omelchenko_2013,Panaggio2015}.

The existence of chimera states challenge the assumption that symmetric systems always yield symmetric outcomes, and points out the subtle mechanisms involve in pattern formation and symmetry breaking in dynamical networks. Their study has implications across disciplines, including biology, reaction-diffusion systems, quantum spin chains, power grid dynamics, and even social coordination\cite{Bansal2019, Li2024, Sakurai2021, Pikovsky2021}, making them a powerful concept in understanding the richness of behavior in complex systems \cite{Panaggio_2015}. 

Recently, this symmetry-breaking physical concept associated to the emergence of chimeras states has attracted great attention in neuroscience since its presence has been related, e.g., in unihemispheric sleep in birds and dolphins where one cerebral hemisphere shows high‐amplitude, synchronized slow EEG waves while the other remains in a low-amplitude, desynchronized, wake-like state \cite{Rattenborg2000, Glaze2021}. Following these findings, many recent studies have been trying to determine the biological mechanisms that give rise to coexisting coherent and incoherent population activity states in different neural systems \cite{TSIGKRIDESMEDT2015,CALIM2020108,santos2019,provata2019}.

Complex patterns in biological, chemical, and atmospheric systems typically arise only under conditions far from equilibrium. Such dissipative structures result from continuous, non-balanced exchanges of energy and matter with the environment and are maintained by fluxes imposed by external forces or constraints. Energy gradients drive irreversible processes that generate and preserve macroscopic order while exporting entropy to the surroundings. In the absence of external driving, these structures decay, and the system relaxes to its equilibrium free-energy minimum \cite{Prigogine1984-PRIOOO-2, Tranquillo2019}.

As usually described in the literature \cite{abramsstrogatz2004,
Panaggio2015,Omelchenko2018}, chimeras as stable attractors cannot be described by a static hamiltonian measure. Detailed balance is violated, confirming that they appear in far-from-equilibrium regimes.  Variants of the initial formulation of the Kuramoto model \cite{Kuramoto1975SelfentrainmentOA}, for instance, with cosine global coupling are fully integrable and a Hamiltonian function can describe the dynamic but no chimera states appear \cite{Watanabe1993}. It is only when we introduce a phase lag between oscillators that these structures appear \cite{Panaggio2015}. On the other hand, early studies generally  assumed that dissipation was essential for the appearance of chimera states. However, a recent work \cite{Witthaut2014} that focused on a Hamiltonian formulation of oscillations and the emergence of synchrony proved the existence of Kuramoto dynamics \cite{Kuramoto1975SelfentrainmentOA} in a Hamiltonian system, for a class of its invariant tori, thus, distinctly linking dissipative to conservative dynamics and suggesting that chimera states may also occur in Hamiltonian systems. 
More recently, it has been shown that chimeras can also arise in quantum conservative Hamiltonian systems with nonlocal hopping, where the energy is well-defined and conserved \cite{Lau2023}. This work proposes the construction of a multi-component Bose–Einstein condensate (BEC) with nonlocal coupling that admits chimera solutions and that could be experimentally realizable \cite{Lau2023}. Nevertheless, a rigorous proof that chimera states exist in conservative systems has not been developed yet.
On the other hand, it has been reported that chimera states are not, in fact, stable attractor states but rather long-lived solutions or quasi-stationary structures \cite{chimeratransient2011}.

A paradigmatic framework for studying complex emergent phenomena arising from the cooperative interactions among system elements -- such as chimera states -- is the Ising model \cite{Ising1925,onsager1944}. This model, together with its variants, has been extensively applied to investigate magnetic phenomena -- including spin glasses~\cite{edwards1975} -- lattice gases, opinion dynamics, social contagion, and neural networks -- both artificial and biological \cite{hopfield1982}. The Ising model has also influenced the development of classical machine learning methods, such as Boltzmann machines \cite{hinton1985}. Fundamentally, it describes a system of binary variables governed by a Hamiltonian formulation, which undergoes a transition from disorder to order as the temperature decreases.

Although Ising-like systems have been widely employed to study diverse physical and biologically inspired phenomena, only a single work has so far reported the emergence of a chimera state in an Ising-type spin system \cite{Singh2011}.
In this work it has been constructed a globally coupled model of two equal‐sized spin modules, each module featuring ferromagnetic intra‐module coupling and antiferromagnetic inter‐module coupling under a uniform external field. The authors demonstrated analytically and via Monte Carlo simulations that, over a finite temperature range, one module can become highly magnetized (ordered) while the other remains weakly magnetized constituting a discrete analog of chimera ordering. However, the explicit breaking of homogeneity through distinct coupling signs from the canonical chimera criterion of spontaneous symmetry breaking in an otherwise uniform network.

In the present work, we revisit chimera phenomena in binary-state dynamics by introducing an extended one-dimensional Ising-type model that preserves interaction symmetry and excludes external influences. Our framework is based on a homogeneous network topology with long-range coupling kernels, designed to enable different regimes or phase coexistence without imposing heterogeneity. Within this setting, we demonstrate the emergence of chimera-like patterns -- coexisting regions of static magnetization and rapidly fluctuating spins -- in fully symmetric binary systems. We further identify the parameter regimes supporting stable phase coexistence, investigate the finite-size scaling of chimera domains, and analyze the influence of initial conditions on pattern selection.

Our findings expand the scope of chimera research by demonstrating that even the simplest binary-state frameworks can sustain hybrid, partially ordered states. This opens new avenues for modeling emergent partial synchronization in neural networks, opinion dynamics, and sociophysical systems where binary logic dominates.

The present work is organized as follows. Section \ref{models} introduces the mathematical model under study, focusing on a non-local purely diffusive process in an Ising spin chain, and describes the Monte Carlo algorithm used for simulations.  In Section \ref{analyt}, we employ an inductive analytical procedure to characterize key behaviors of the system. Section \ref{comp} then presents numerical simulations of the system,  alongside a comparison between theoretical predictions and  numerical results.

Our study shows that a simple Ising chain with non-local diffusion  exhibits a rich phenomenology,  including chimera states. In this scenario, we also show that these states constitute the  equilibrium state satisfying detailed balance, for certain regions of the parameter space. The present work could be extended to some non-equilibrium situations, including, e.g., competing reaction and diffusion dynamics, a scenario more commonly observed in many social, ecological and biological systems. 
{In this last case, a prominent example is the activity of the mammalian brain, where neural excitability is driven by chemical synapses (reaction) and voltage propagation occurs via gap junctions or electrical synapses among neurons (diffusion) \cite{Koch1998,Synapses3,Synapses2}. Then our model can be easily extended to describe brain activity at multiple scales. At the microscopic level, spin variables represent neuronal voltage states, firing or not, with diffusion linked to the presence of electrical synapses in the neural tissue. At mesoscopic or macroscopic scales, it could be used to study local field potential diffusion between neighboring brain regions \cite{Synapses3,Synapses1}}.

\section{Models and Methods}\label{models}

\subsection{Non-local diffusion dynamics}\label{sec:dif}
\begin{figure}[t]
\begin{center}
\includegraphics[width=\linewidth]{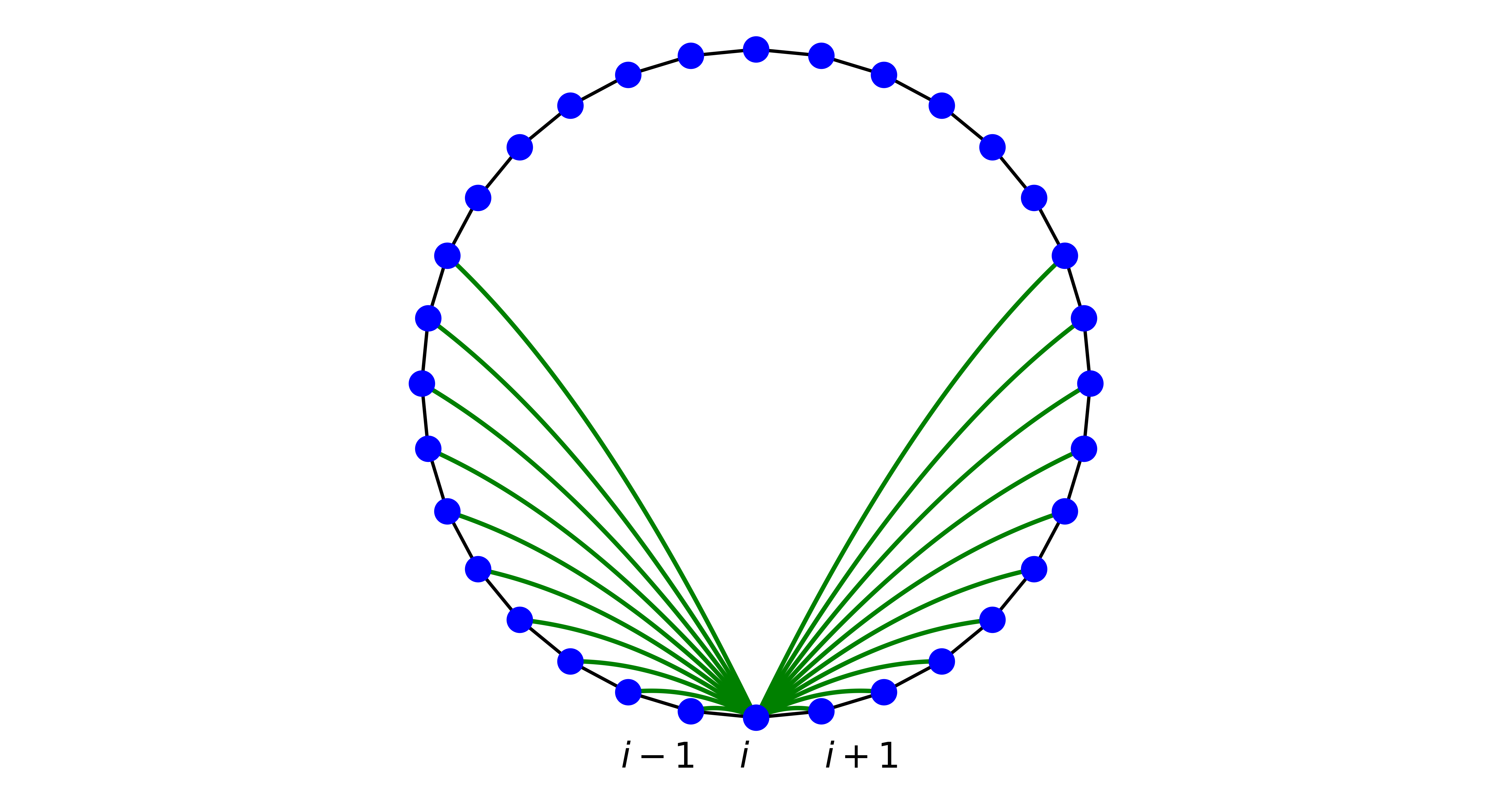}
\end{center}
    \caption{{\bf System connectivity scheme used in the present work:} Schematic of a one-dimensional ring with $N=30$ spins showing long-range Kawasaki interactions. Each spin (blue circle) connects to $R=10$ neighbors on each side (green lines), enabling diffusion through spin-exchange dynamics. The periodic boundary conditions create a closed ring topology.}
    \label{fig:circlekawa}
\end{figure}

In our study, we consider the case of diffusion dynamics on a one-dimensional lattice of $N$ Ising spins. The state of the system is represented by $\boldsymbol{\sigma}=\{\sigma_1,\sigma_2, ... , \sigma_N\}$, where $\sigma_i=\{-1,1 \}.$ We consider also periodic boundary conditions $\sigma_{N+1}=\sigma_1$ (see Fig. \ref{fig:circlekawa}). With such a spin ring in contact with a thermal bath at temperature $T$, we assume that spins relax to equilibrium, conserving the total magnetization through spin exchange or Kawasaki dynamics \cite{Kawasaki1966} with an effective Hamiltonian $\mathcal{H}(\boldsymbol{\sigma})$:

\begin{equation}
    \mathcal{H}(\boldsymbol{\sigma})=-\frac{1}{2}\sum_{i,j} J_{ij}\sigma_i\sigma_j 
    \label{eq:hamiltoniank}
\end{equation}

where 
$J_{ij}=J\times[0<|i-j|\leq R]$ (with $[P]$ indicating the Iverson bracket) indicates constant ferromagnetic interaction strength among spins ($J>0$) within the diffusion range $R$. {Note that ferromagnetic interaction strength $J$ among spins} is given in units of $k_{B}T.$ An illustration of the system is depicted in Fig. \ref{fig:circlekawa}.

{Due to the presence of the thermal bath, the system evolves according to stochastic dynamics governed by a master equation -- with transition matrix $W(\boldsymbol{\sigma}^\prime,\boldsymbol{\sigma})$ -- 
which gives the evolution in a characteristic time $\tau$ of the probability $P(\boldsymbol{\sigma},t)$ of finding the spin system in state $\boldsymbol{\sigma}$ at time $t$ \cite{Mastereq2, Mastereq1, Mastereq3}, and that can be expressed as} 
\begin{gather}
\tau\frac{dP(\boldsymbol{\sigma},t)}{dt}=\sum \limits_{\boldsymbol{\sigma}'}[P(\boldsymbol{\sigma} ',t)W(\boldsymbol{\sigma} ', \boldsymbol{\sigma})-P(\boldsymbol{\sigma},t)W(\boldsymbol{\sigma},\boldsymbol{\sigma} ')]
     \\ W(\boldsymbol{\sigma} ',\boldsymbol{\sigma})=\sum\limits_{\substack{i,j \\ 0<|i-j|\leq R}} \delta_{\sigma_1,\sigma'_1}...\delta_{\sigma_i\sigma'_{j}}...\delta_{\sigma_j\sigma'_{i}}...\delta_{\sigma_N,\sigma'_N}w^{ij}(\boldsymbol{\sigma})
     \label{masterdif}
\end{gather}
with spin exchange transition probabilities per unit of time $w^{ij}(\boldsymbol{\sigma})$ ($\sigma_i\longleftrightarrow \sigma_j$) given by the Metropolis \cite{metropolis} algorithm:

\begin{equation}
    w^{ij}(\boldsymbol{\sigma})=\begin{cases}
        1 \quad \mathrm{if }\; \Delta \mathcal{H}^{ij}\leq 0\\
        e^{-\beta \Delta \mathcal{H}^{ij}} \quad \mathrm{if }\; \Delta \mathcal{H}^{ij}> 0 \
    \end{cases}\quad\quad \Delta \mathcal{H}^{ij}=
        \mathcal{H}(\boldsymbol{\sigma}^{ij})-\mathcal{H}(\boldsymbol{\sigma}) 
    \label{eq:metropolis}
\end{equation}
where the spin configuration $\boldsymbol{\sigma}^{ij}$ has the spin state variables at $i$ and $j$ exchanged compared with configuration $\boldsymbol{\sigma}$ and moreover $|i-j|\le R.$
The detailed balance condition, which states $P(\boldsymbol{\sigma} ',t)W(\boldsymbol{\sigma} ', \boldsymbol{\sigma})=P(\boldsymbol{\sigma},t)W(\boldsymbol{\sigma},\boldsymbol{\sigma} ')$, is then fulfilled. This means that there is a stationary state and that there is no entropy production in that state. Starting at a random configuration, the system then relaxes to equilibrium through a non-local diffusion process.

\subsection{Simulations and numerical details}\label{algorithm}
We now describe the Monte Carlo algorithm used to simulate our system for an extensive combination of model parameters. The Ising model itself has no intrinsic dynamic and it is when we couple the system to a heat bath at temperature $T$ that stochastic changes in the spin variables occurs so we have a stochastic kinetic version of Ising model driven by the master equation as those in Eq. (\ref{masterdif}). In a spin system of size $N$, each Monte Carlo Step (MCS) consists of $N$ spin-flip or spin exchange attempts. In general, we have performed simulations of $5\times 10^4$ MCS. Each simulation of our system operates according to the following algorithm:

\begin{enumerate}
 \item First, two integers $i,j$ are randomly picked from $U\{1,2,...,N\}$ . We then check whether these indices are connected ($J_{ij}>0$) and we start a new try if they are not. \label{step3}
 
\item The Metropolis rate $w^{ij}(\boldsymbol{\sigma})$ is then evaluated following the algorithm described in Eq. (\ref{eq:metropolis}) and we compute $\Delta \mathcal{H}$ explicitly as follows: 

\begin{equation}
    \Delta \mathcal{H}^{ij}=\left(\sum\limits_{\substack{\\ k\neq j} }J_{ik}\sigma_{k}-\sum\limits_{\substack{ k'\neq i} }J_{jk'}\sigma_{k'}\right) \left(\sigma_i -\sigma_j\right) \label{deltakawa}
\end{equation}

 \item A random number $\zeta$ is generated from $U(0,1)$. If $\zeta<w^{ij}(\boldsymbol{\sigma})$, the spins $i$ and $j$ are exchanged.   

\end{enumerate}
{Spin pairs in the ring are updated using a sequential updating, i.e., at each time step during the evolution of the system one chooses a spin pair at random and exchanges their state using the Metropolis algorithm. Then the system state is updated after each attempt. Note that this is different to what occurs during parallel updating in which, for example in the case of spin-flip dynamics, one performs $N$ spin-flip attempts and updates the state of the whole system state only after these $N$ attempts.} Although work regarding parallel updating in the Ising model has been conducted \cite{KUMAR2024129773} and proven to obey equilibrium statistics, one has to be careful with its implementation so that it satisfies detailed balance \cite{Nilmeier2014}. 

The simulations presented in our study were performed using C++, resulting in a significant improvement in both simulation performance and the computational cost of numerical analysis. We were able to parallelize different runs with \textit{OpenMC} and the use of efficient random number generators as \textit{mt19937$\_$64} made large system sizes fast to compute. Nevertheless, the use of \textit{Python} was necessary in the integration of different parts of the code and in the visualization of data. 
\section{Results}

\subsection{Analytical results}\label{analyt}
The difference of energy for the exchange of two spins chosen at random, namely $\sigma_i$ and $\sigma_j,$ is:

\begin{equation}
    \Delta \mathcal{H}^{ij}=\left(\sum\limits_{\substack{k=i-R \\ k\neq i,j} }^{i+R}\sigma_{k}-\sum\limits_{\substack{k'=j-R \\ k'\neq i,j } }^{j+R}\sigma_{k'}\right) \left(\sigma_i -\sigma_j\right) 
    \label{energy}
\end{equation}
where we have just computed the terms in $\Delta\mathcal{H}$ summing over $R$ neighbors on each side and considered $J=1$ for simplicity. Note that exchange energy $\Delta\mathcal{H}^{ij}$ is only non-zero if the exchanged spins $i$ and $j$ have different sign. 

For simplicity, we are going to consider moreover the zero temperature case ($T=0$), where the Metropolis acceptance probability is $1$ for $\Delta \mathcal{H}\leq 0$ and $0$ for $\Delta \mathcal{H}>0$. This zero temperature choice will allow us to obtain some analytical derivations and conclusions concerning the appearance and stability of chimera states.

\subsubsection{Emergence and stability of chimera states }\label{chimerastates}
We investigate wether chimera states can emerge in a ring of $N$ spins under non-local diffusion.
In principle, one can define for example a chimera state in this ring
as a set $n^{+}$ of positive spins located at random ordered positions
$i_{1}<i_{2}<\ldots<i_{n^{+}}$ that can exchange each other via
Kawasaki dynamics. If the condition $|i_{1}-i_{n^{+}}|<R$ is met, all different-signed spin can be exchanged. This configuration is  analogous
to the well-known chimera states in a ring of Kuramoto oscillators
since now there is a subpopulation of spins with alternating values (which we can consider equivalent to an unsynchronized population)
and other subpopulation in which all spins are in the same state (equivalent
to a synchronized population).  Moreover, since the Ising model is symmetric under global spin flip, any result we derive for
a given number $n^+$ of positive spins in an otherwise negative background will have an equivalent
interpretation for $n^{-}$ negative spins in an otherwise positive background.
 
At each time step of the evolution, if one has a stable chimera domain,
it is convenient to define the set of indices which can be exchanged
with the $n^{+}$ positive spins within the chimera domain as
\begin{equation}
B_{n^{+}}(i_{1},\ldots,i_{n^{+}})=\left\{ k\in\left\{ 1,\ldots,N\right\} ,\;k\in\bigcap_{m=1}^{n^{+}}B_{1}(i_{m}),\; |i_1-i_{n+}|\leq R\right\} \label{balln}
\end{equation}

with 
\[
B_{1}(i_{m})=\left\{ k\in\left\{ 1,\ldots,N\right\} ,\;|k-i_{m}|_{p}<R,\;k\neq i_{m}\right\} \quad \forall m=1\ldots n^{+}
\]
and where $|k-i_m|_p\coloneqq min\{|k-i_m|,\,N-|k-i_m|\}$ to account for the existence of periodic boundary conditions in the ring. Note that this definition of distance is general and can be extended for any pair of sites ($k,l$). Moreover, if $B_{n^{+}}(i_{1},\ldots,i_{n^{+}})$ is not empty, by definition it only contains negative spins, which can be exchanged
with any of the positive spins $\sigma_{i_{1}},\;\sigma_{i_{2}}\,... \,\sigma_{i_{n^{+}}}$.

Given two indices chosen at random we can compute the transition rates $W(\boldsymbol{\sigma}',\boldsymbol{\sigma})$ by considering their exchange. Only transitions of different-signed spins change the configuration, and we say that this happens with probability   $p_{n^+}^{exchange}$. On the other hand transitions that exchange negative or positive spins will not modify the configuration and this will occur with probability of ``copy'' the initial configuration, namely $p_{n^+}^c=1-p_{n_+}^{exchange}.$ One can easily compute then exchange and copy probabilities as:
\begin{equation}
p_{n^+}^{exchange}=\frac{2n^+}{N^2} | B_{n^+}(i_1,...\,,i_{n^+})|\;\;\;\;\;\;\;\;\;\;\;\; p_{n^+}^{c}=1-p_{n^+}^{exchange}.
\end{equation}

We can now construct a general transition probability for any pair of randomly (and independently chosen)  unordered indices  $(i,j)$ using the exchange operator
\[
  Ex_{\,i,\,j}(\boldsymbol{\sigma} ',\boldsymbol{\sigma}) 
  \;:=\;
  \delta_{\sigma'_{i},\,\sigma_j}\,
  \delta_{\sigma'_j,\,\sigma_{i}}\,
  \prod_{\,k \neq i,\,j}\,
    \delta_{\sigma'_k,\,\sigma_k}=\begin{cases}
        1\; \text{ if $\sigma'$ equals  $\sigma$ exchanging spins at $i $, $j$}\\
        0 \;\text{ otherwise}
    \end{cases}
\]
and the configuration Kronecker delta operator defined as
\[
  \Delta(\boldsymbol{\sigma} ',\boldsymbol{\sigma}) 
  \;:=\;
  \prod_{\,k}\,
    \delta_{\sigma'_k,\,\sigma_k}=\begin{cases}
        1\; \text{ if $\boldsymbol{\sigma}'$ =  $\boldsymbol{\sigma}$ }\\
        0 \;\text{ otherwise}
    \end{cases}
\]

as:
\begin{equation}\label{eq:Kawasakin}
W\bigl(\boldsymbol{\sigma}',\boldsymbol{\sigma}\bigr)
\;=\;\frac{p_{n^+}^{exchange}}{n^+}\sum\limits_{k=1}^{n^+}\left[\sum\limits_{i_k,j\in B_{n^+}(i_1,... ,i_{n^+})}Ex_{i_{k},j}(\boldsymbol{\sigma}',\boldsymbol{\sigma}\bigr)\right]+p_{n^+}^c\;\Delta(\boldsymbol{\sigma}',\boldsymbol{\sigma}\bigr)
\end{equation}

If we denote by  $\Lambda\equiv\mathbb{Z}_{2}^{N}$ the set of system
configurations, we can define the following subset of system configurations
\begin{equation}
\varSigma^{n^{+}}:=\left\{ \boldsymbol{\sigma}\in\Lambda\;/\;|i_{1}-i_{n^{+}}|_{p}\leq R,\;i_{1}<i_{2}<\ldots<i_{n^{+}}\right\}. \label{chimera}
\end{equation}

The set $\varSigma^{n^{+}}$ defines then the general class of configurations that support chimera states. In fact, all transitions between  system configurations in $\Sigma^{n^+}$, which is the energy minimum, are equally likely. These correspond to a ring configuration with a noisy region, which we refer to as \emph{chimera region}. It will be of size 
\begin{equation}L_{chimera}\coloneqq |B_{n^+}(i_1,...\,,i_{n^+})|+n^+
\label{lengthchimera}
\end{equation}
where we have adopted the convention to include the indices to be exchanged to the \textit{chimera region}. This region is stable in time, since any of these configurations is already at an energy minimum, and therefore the Metropolis transition probability forbids any transition in which $\Delta \mathcal{H}\ne0$, that is any non-trivial  transition that involve indices outside $L_{chimera}$.


We now consider the case in which we do not have a configuration of type $\Sigma^{n^+}$ from the start, but that we have the $n^{+}$ positive spins randomly
distributed over the ring. Suppose that the system is at a state in its evolution
such that two positive spins located at sites $i_{n}$ and $i_{m}$
can not interact, i.e. $|i_{n}-i_{m}|_{p}>R$, but at least a negative
spin located at $k\in B_{1}(i_{n})$ could be exchanged with $i_{m}$, which would be a negative-energy transition. As result the system will change to a configuration
such as now $|i_{n}-i_{m}|_{p}\leq R.$ Even if $k\notin B_{1}(i_{n})$, the fact that stochastic nature of proposed exchanges and the fact that non-positive energy exchanges are accepted ensures that eventually all spin pairs will arrive to a configuration in which $|i_{m}-i_{n}|_{p}\leq R$.

These negative-energy
changes will occur until the system reaches a configuration where
the distance between all positive spins is less than $R$ which
will correspond to a configuration $\boldsymbol{\sigma}\in\varSigma^{n^{+}}$,  being the energy absolute minimum of the dynamics.

Also, note that the actual construction of $B_{n^+}(i_1,...\,,i_{n^+})$ is not always possible, as  can be see from definition (\ref{balln}). This follows from the fact that the interaction range $R$ limits the maximum number of positive spins that can simultaneously located within the range of another one. In fact, for configurations with $n^+> R+1 \Rightarrow |i_1 -i_{n^+}|_p > R \Rightarrow \; B_{n^+}=\emptyset$. In this situation, there is no rule to determine which spins will have zero energy transitions available. Actually, even for a configuration with $n^+=R+1$, regardless of whether $|i_1- i_{n^+}|_p\le R$, the ball would still be the empty set. This is because any individual ball $B(i_k)$ (from which we will construct $B_{n^+}$) is defined without the index $i_k$. This does not imply that exchanges are impossible, but that they are trivial as all spins are positive. We will try to discuss briefly what happens for $n^+>R+1$ in the next section.

\subsubsection{Disappearance of chimera states and formation of attractors} \label{attractorsec}
We have seen above that  whenever the number of positive spins exceeds \(R\), any non‐trivial chimera region must shrink and eventually disappear.  In particular, let \(n^+ = R+1\) and consider the following set of ordered positive spin indices 
\begin{equation*}
   \pi\coloneqq \{i_1 < i_2 < \cdots < i_{R+1}\}
\end{equation*}
If $i_1$ and \(i_{R+1}\) are at distance \(R\)  the condition $|i_1- i_{R+1}|_p\le R$ holds. This would be, for instance, a configuration of the type
\[
\boldsymbol{\sigma}=\{\sigma_1^-,\,...,\sigma_{i-1}^-,\sigma_i^+,\sigma_{i+1}^+,... ,\sigma_{i+R+1}^+,\sigma_{i+R+2}^-,\,...\,\sigma^-_N\}
\]
where we employ the notation $\sigma_i^{\pm}$ to denote positive/negative spins at site $i$. In this case, then one has $i_1\in B_1(i_{R+1})$ and since in general $i_k\notin B_1(i_k)$ by definition, it follows $B_{\,R+1}\bigl(i_1,\dots,i_{R+1}\bigr)\;=\;\emptyset$. This configuration would be a rather extreme case, since it fulfills the condition to be in $\Sigma^{R}$, and in principle $B_{R+1}$ is realizable, although the exclusion of the centers of the balls ($i_k\notin B( i_k)$) make it to be the empty set. 

Nonetheless, we have arrived to the energy minimum ($\boldsymbol{\sigma} \in \Sigma^R$), though any possible exchange would not change the configuration. Following equation (\ref{eq:Kawasakin}):
\begin{equation*}
W(\boldsymbol{\sigma}',\boldsymbol{\sigma}\bigr)
\;=\; \Delta(\boldsymbol{\sigma}',\boldsymbol{\sigma}\bigr).
\end{equation*}
In this configuration, taking into account that every positive spin is inside every other positive spin's ball ($|i_n^+-i_m^+|_p\le R  \;\forall\;n,m$), every exchange would be between two positive spins, creating a copy of the system. Then, the noise characterizing the \emph{chimera region} has disappeared.

Let us now extend the number of positive spins considering now $n^+>R+1$. In this situation, $B_{n^+}=\emptyset$ as in the last case, but an important difference is that configurations in the energy minima will not be a configuration of type $\boldsymbol{\sigma}\in\Sigma^{n^+}$, since for $n^+>R+1$ one has $|i_1 -i_{n^+}|_p>R$, so no configuration satisfies the condition to be in $\Sigma^{n^+}$, from definition (\ref{chimera}). The system will anyway advance stochastically to an energy minimum, but this minimum will not be in any class of type $\Sigma^X$. However, we can give an argument  of how these minima behave. 

\theoremstyle{proposition}
\newtheorem{proposition}{Proposition}[subsection]

\begin{proposition}[Energetic asymmetry of boundary spins]\label{prop}
Let \(R>1\) be fixed.  Suppose 
\[
n^+ > R+1,
\]
and let 
\[
\pi \;=\;\bigl\{\,i_1<i_2<\cdots<i_{n^+}\bigr\}
\]
be the ordered indices of all \(n^+\) positive spins on the ring.  Assume moreover that each consecutive pair satisfies $|i_{k}-1- i_{k+1}|_p\le R\;\;\forall k\in\pi$ excluded the first and the last,  so that each positive spin is within the exchange range \(R\) of its immediate neighbors in \(\pi\). Let us denote by $\Pi$ the set of ring spin configurations satisfying these requirements. Then, at zero temperature:

\begin{enumerate}
  \item The leftmost positive spin \(i_1\) has at least one exchange available with a neighboring negative spin $k$ in the set
  \[
    B^-(i_1)\coloneqq B_1(i_1)\,\cap\,B_1(i_2)\,\cap\,B_1(i_3)
  \]
  such that the exchange \(\sigma_{i_1}\leftrightarrow\sigma_{k}\in B^-(i_1)\) strictly \emph{decreases} energy, whereas every exchange of \(i_1\) with a site $k'$ in
  \[
  B^+(i_{1})\coloneqq B_1(i_1)\setminus B_1(i_2)
  \]
where $B^+(i_{1})$ contains all indices in $B_1(i_1)$ that are not in $B_1(i_2)$, strictly increases energy, and is therefore forbidden.

 \item Similarly, the rightmost positive spin \(i_{n+}\) has at least one exchange available with a neighboring negative spin $k$ in
  \[
    B^-(i_{n^+})\coloneqq B_1(i_{n^+})\,\cap\,B_1(i_{n^+-1})\,\cap\,B_1(i_{n^+-2})
  \]
  such that the exchange \(\sigma_{i_{n^+}}\leftrightarrow\sigma_{k}\in B^-(i_{n^+})\) strictly \emph{decreases} energy, whereas every exchange of \(i_{n^+}\) with a site $k'$ in
  \[
  B^+(i_{n^+})\coloneqq B_1(i_{n^+})\setminus B_1(i_{n^+-1})
  \]
  strictly increases energy, and is therefore forbidden.

\end{enumerate}
\end{proposition}

\begin{proof}
Because $|i_1 - 1 - i_2|_p\le R$ and $|i_2 - 1 -i_3|\le R$, we can construct a non-empty ball
\[
B^-(i_1)\neq\emptyset.
\]
Choosing any negative index $k\in B^-(i_1)$ and exchanging the spin with $i_1$,  
\(i_1\leftrightarrow k\) moves a positive spin into a site that remains within \(R\) of both \(i_2\) and \(i_3\).  Thus \(i_1\) gains at least one extra positive neighbor, strictly lowering the energy.  In contrast, if 
\(\ell\in B^+(i_{1})\ne\emptyset\), 
then \(\ell\) is outside \(B_1(i_2)\) and therefore cannot lie within \(R\) of any other positive. Exchanging \(i_1\) with \(\ell\) implies to lose its only positive neighbor \(i_2\) without gaining any, strictly raising the energy.  That shows part 1.

For the rightmost spin \(i_{n^+}\), the same argument applies: since $|i_{n^+-1} - 1 -i_{n^+}|_p\le R$ and $|i_{n^+-2} - 1 - i_{n^+-1}|_p\le R$  we can construct a non-empty ball 
\[
 B^-(i_{n^+})\neq\emptyset.
\]
Any $k'\in B^-(i_{n^+})$ satisfies \(\sigma_{k'}=-1\) and lies within \(R\) of two other positives, so swapping \(i_{n^+}\leftrightarrow k'\) strictly lowers the energy.  On the other hand, any 
$\ell'\in B^+(i_{n^+})\ne \emptyset $ lies outside $B_1(i_{n+-1})$ and  cannot lie within $R$ of any other positive. Then an exchange of $\ell'$ with $i_{n^+}$ would not be allowed, since it will increase the energy due to the lost of its only positive neighbor $i_{n^+-1}$.  This proves part 2.
\end{proof}

Consequently, if a boundary spin is selected for an exchange attempt, it is forced to move ``inward'' toward the center of the configuration  $\sigma\in\Pi$ . More precisely, it is prevented from moving ``outward'' the positive block, and given that all sites are equally probable for an exchange attempt, the resulting behavior is that it is forced to move ``inward''.  Each such move strictly reduces the total span of $\pi$, that is, the periodic distance $|i_{n^+}-i_1|_p$. Repeating these energy‐lowering boundary spin exchanges at \(T=0\) forces the positive cluster to shrink until its length is \(n^+\).  As a consequence, we conclude that no $\emph{chimera state}$ can exist whenever $n^+>R+1$.

To summarize, the asymmetry in energy connections for the first and last positive indices in  $\sigma\in\Pi$ causes the domain to reduce its length, thereby decreasing the span of $\pi$ until it reaches $n^+$. Consequently, the configuration evolves toward the global energy minimum; we henceforth refer to such configurations as  \emph{attractors}.

To deduct the formation of these \emph{attractors}, we have supposed that we have a configuration of type $\Pi$ from the start (see proposition \ref{prop}). This will obviously not always be the case, and to characterize any kind of general behavior we have to employ a more general approach. In fact, we realized that the Hamiltonian \eqref{eq:hamiltoniank} is a special case of a one‐dimensional Ising model with Kac‐type interactions \cite{Cassandro1993,presuti1d,presutti}. For these, it is shown that regions of constant magnetization $m$ and $-m$ alternate throughout the system (what we have called \emph{attractors}), and that they are of size $L_{attractor}\approx e^{cR} $ for some suitable constant $c$, which we will numerically calculate in the next section.

\subsection{Numerical results for T=0}\label{comp}
\begin{figure}[ht!]
\begin{center}
\includegraphics[width=\linewidth]{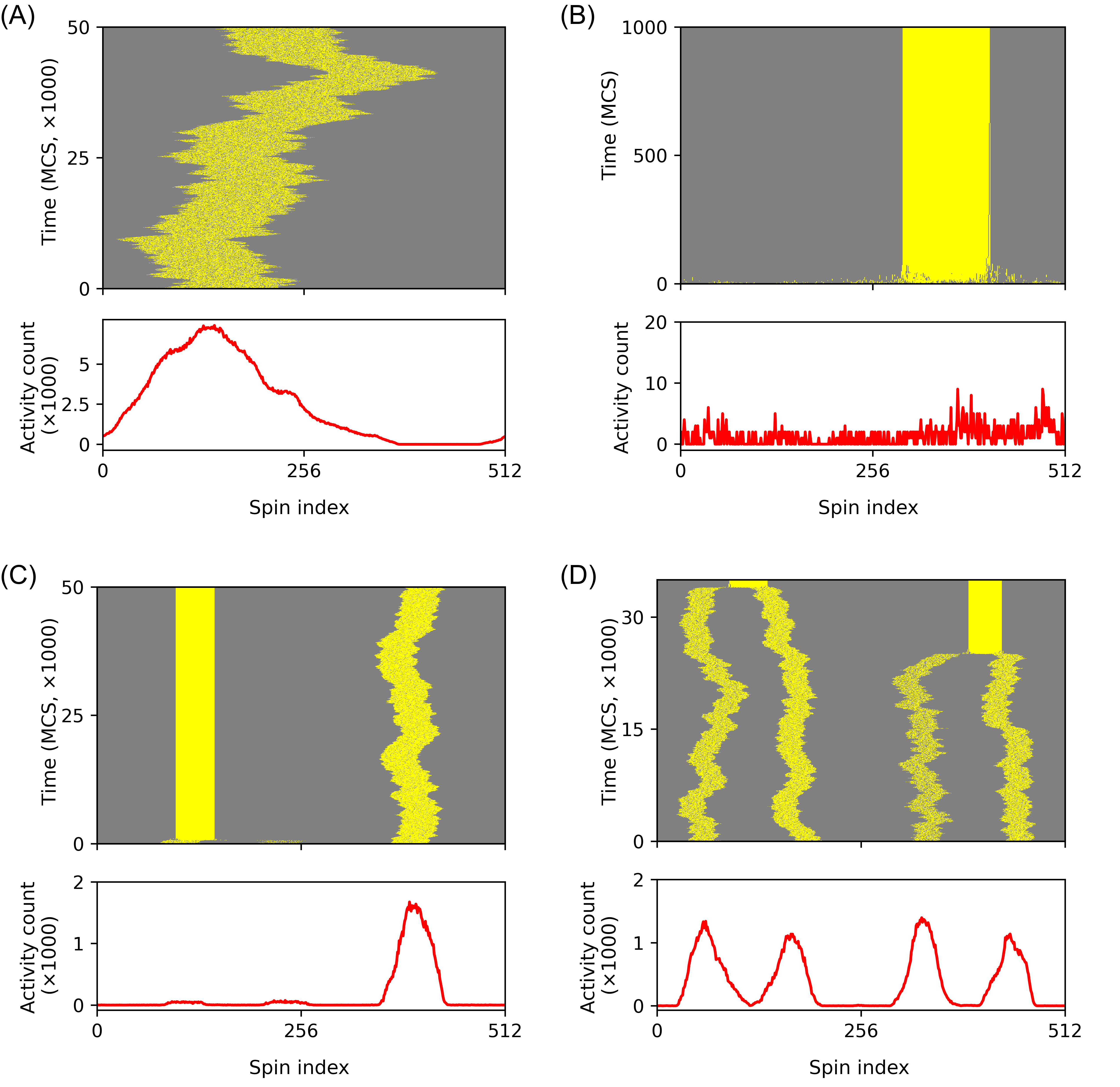}
\end{center}
\caption{\textbf{Four observed emergent behaviors in the system for a $N=512$ spins ring:} (A) chimera states, (B) attractor states, (C) coexistence of chimeras and attractors, and (D) merging of chimeras into attractors. From top to bottom and from left to right model parameters were $r=0.264, 0.205, 0.088, 0.059$ and $|m_0|=0.65,0.55,0.65,0.65,$ respectively. }
\label{types}
\end{figure}

We simulated a system of $N=512$ Ising spins over a maximum time of $t_{max}=5 \cdot 10^4$ MCS in the zero temperature case, as described in Section \ref{algorithm}. Using the averaged magnetization of a system with $N$ spins in which all are negative except $n^+$ positive spins
\[
 m_0 = \frac{2n^+}{N} - 1,
\]
we explored the emergent features of the system for various values of the normalized interaction range $r=R/N$ and $m_0$. We obtained different types of behavior, as illustrated in Fig. \ref{types}. 
We illustrate in the top panels the time series of the state of the ring in the form of raster plots. Each site $i$ is depicted in gray if $\sigma_i=-1$ and in yellow if $\sigma_i=1$. The evolution of the configuration $\boldsymbol{\sigma}$ is shown by observing the evolution of spins $\sigma_i$. 
Also, below each raster plot we have represented the \emph{activity} of each site throughout the  duration of the simulation,  which helps us  distinguish in quantitatively between different behaviors and to visualize the displacement of chimera states. Here, activity is defined as the number of times a particular spin $\sigma_i$ changes sign during the simulation. The index of each spin is displaced to account for the displacement of the region in which that spin is through the  density-based clustering algorithm \textit{DBSCAN} \cite{Ester1996ADA}. We found four different type of behavior, namely:
\begin{enumerate}
    \item \textbf{Chimera state phase} (see Fig. \ref{types}(A): In line with derivations in Section~\ref{chimerastates}, we observe in this case one chimera state, constituted by a domain of desynchronized spins coexisting with a constant or synchronized local magnetization domain. Note that a main feature of emerging chimeras in the Ising chain is that the domain of desynchronized spins changes in time. This fluctuating behavior decreases when $N$ increases while keeping $r=R/N$ constant (see top panel of Fig. \ref{sizecomparison}).
    \item \textbf{Attractors phase} (Fig.~\ref{types}(B)): These correspond to coexistence of fully segregated positive and negative domains of constant local magnetization, as predicted in Section~\ref{attractorsec}.
    \item \textbf{Coexistence of chimeras and attractors} (Fig.~\ref{types}(C)): This corresponds to one unexpected result, where  emerging chimera states coexist with attractors.
    \item \textbf{Transient chimeras phase} (Fig.~\ref{types}(D)): In this phase another unforeseen result emerges characterized by existence of transient chimeras which, after some time, collapse into stable attractors.
\end{enumerate}

While the attractor and pure chimeric behaviors are fully explained by the analytical framework in Section \ref{analyt}, the coexistence of chimeras and attractors and the merging of chimeras into attractors highlights new kind of states and transitions that could not be analytically predicted using our previous oversimplified analytical   approach. However, we will derive, analyze and test below a plausible dynamics based in simple considerations for the evolution of the number of chimeras and attractors in such intriguing phase. Remembering the necessary condition for the formation of attractors, i.e. $n^+>R+1$ as shown in section \ref{attractorsec}), one can rewrite it in terms of $m_0$ as
\begin{equation}
    m_0>\frac{2(R+1)}{N}-1=2r-1+ {\cal{O}}\left(N^{-1}\right)=m_c
    \label{crit}
\end{equation}
Then, if we see that chimeras and attractors coexist in such conditions, this implies that that chimera state will be \emph{metastable}, as chimeras will always be able to merge into attractors when (\ref{crit}) holds (see also next section).

\subsubsection{Length of attractors}\label{length}
We tested here the analytical prediction of exponential domain–size growth for attractors, $L_{attractor} = e^{cR}$ (at the end of Section~\ref{attractorsec}), by measuring, at the last MCS of each simulation, the \emph{Normalized Mean Domain Length} or $\ell_{dom}$:
\[
\ell_{dom} \;=\; \frac{1}{\#\{\text{domain boundaries}\}},
\]
where a domain boundary is any site for which $\sigma_i \neq \sigma_{i+1}$. This quantity is accurate to measure the mean length of attractors in the ring. In the case of chimeras states, since there many changes of sing of the spins due to the intrinsic noise characterizing the chimera domain, $\ell_{dom}$ is not valid to quantify the number of qualitatively different domains in the ring when only chimeras or coexistence between chimeras and attractors occurs.  

Additionally, we recorded the \emph{Normalized Stationary State Time} or $\tau_{ss}$, that is the MCS step at which $\ell_{dom}$ remains unchanged for $100$ consecutive steps, divided by the total number of steps. This metric reliably indicates the moment at which the system has reached its thermodynamic steady state for pure attractor configurations. Thus, a small $\tau_{ss}$ indicates rapid settling into a stable attractor, whereas a large $\tau_{ss}$ reflects prolonged fluctuations in the number of domain boundaries.
\medskip

We stress that chimera states, as described in Section \ref{chimerastates}, are themselves equilibrium states, whenever $m_0< m_c$. However, recalling that the analytical chimera–formation boundary depends only on $R$ when $m_0$ is fixed, if for a given $R$ chimeras and attractors coexist, the  chimera state must be \emph{metastable}, as we will discuss at the end of Section \ref{mdlsst}.

For both stable and metastable chimera states, the stochastic formation of spin blocks of the same sign within a chimera continuously alters the count of domain boundaries and hence $\ell_{dom}$, so this magnitude never remains strictly constant for the required criterion of 100 consecutive MCS steps without change needed to define $\tau_{ss}$. Therefore we expect higher $\tau_{ss}$ for stable and metastable chimeras states than in purely attractor states, as well as lower $\ell_{dom}$ values, providing a quantitative signature of stable or metastable chimera states.

\begin{figure}[h!]
\begin{center}
\includegraphics[width=0.8\linewidth]{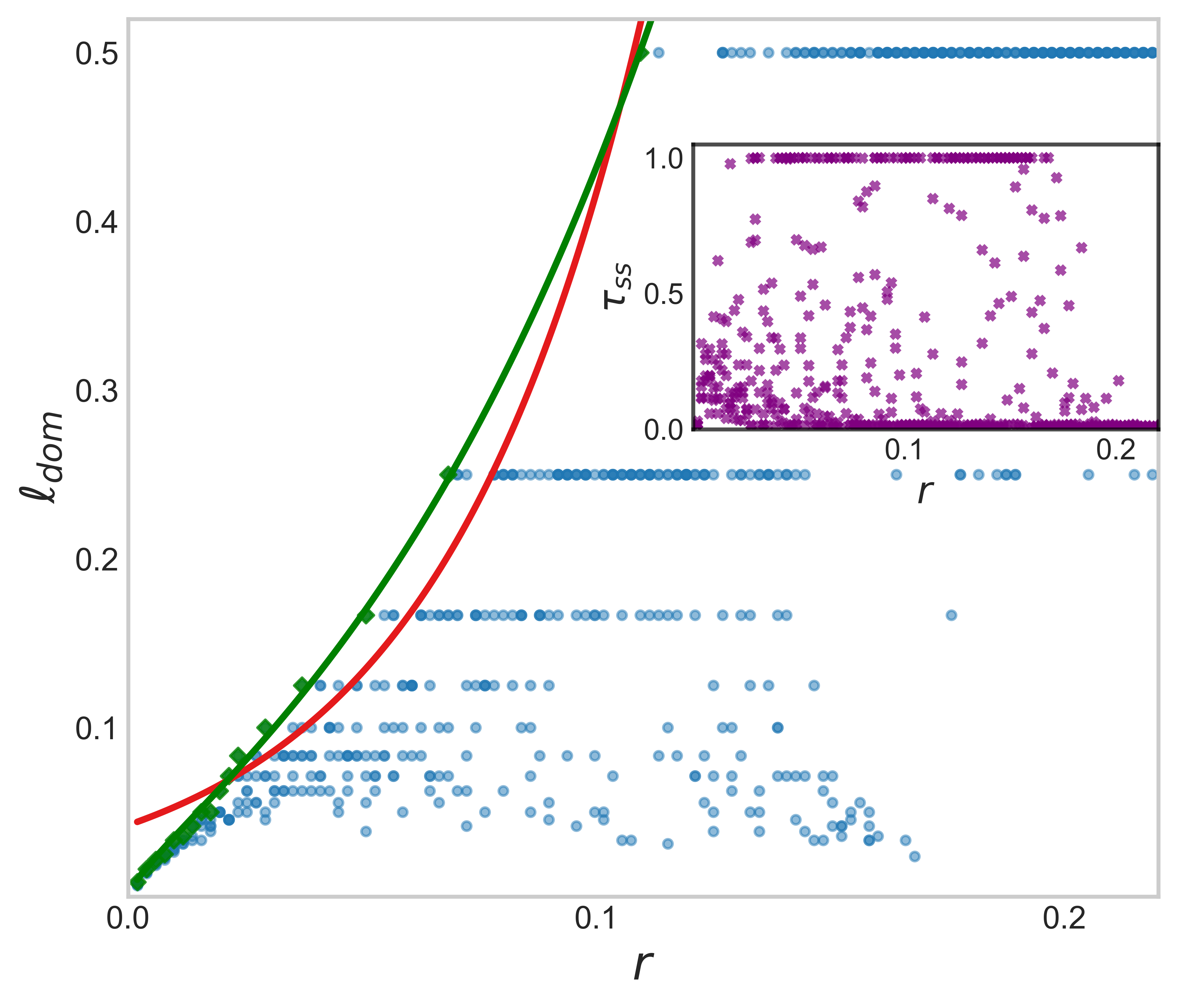}
\end{center}
\caption{Normalized Mean domain Length $\ell_{dom}$ versus interaction range $r=R/N$, demonstrating exponential growth of attractor size with $r.$ Blue circles correspond to all measures of $l_{dom}$ performed in the analysis, while green diamonds correspond to values of $d_{dom}$ used for the fitting. Solid lines correspond to the fit for largest attractor size for a given $r$. In particular the red curve is fitted to  $l^c_{dom}=a\cdot a e^{br}$ and the green curve  corresponds to $l^c_{dom}=x^a\cdot e^{br}.$ The  inset shows the corresponding values of $\tau_{ss}$ for every point in the main plot. Simulations have been performed for $N=512$ and $5\times10^4$ MCS.}
\label{mdl}
\end{figure}

Figure \ref{mdl} plots $\ell_{dom}$ and $\tau_{ss}$ against $r$ for $m_0=0.4$. The figure illustrates that there are multiple $\ell_{dom}$ points for each value of $r$, which correspond to different-seeded runs. The scatter of low $\ell_{dom}$ points across nearly all $r$ values in Fig. \ref{mdl} signals the intermittent presence of small, noise-driven domains, precisely the hallmark of chimera–attractor coexistence seen in Fig. \ref{types}. We fitted $\ell_{dom}$ versus $r$ to two candidate curves:
\[
\ell_{dom}(r) = a\,e^{b r}, 
\qquad
\ell_{dom}(x) = x^a\,e^{b r}.
\]
To select points for these fits, we chose first the \emph{maximum} $\ell_{dom}$ per $r$, thereby excluding runs that could contain chimeras (which would depress $\ell_{dom}$), and secondly, for each observed $\ell_{dom}$ value, we select the \emph{smallest} $r$ that achieved it, ensuring that we captured the minimal interaction range needed to form those domains.

\begin{table}[h]
  \centering
  \begin{tabular}{lcc}
    \toprule
     & $\ell_{dom}(x)=a\,e^{b r}$ & $\ell_{dom}(r)=x^a\,e^{b r}$ \\
    \midrule
    $a$ & $0.042 \pm 0.005$ & $0.743 \pm 0.005$ \\
    $b$ & $22.9  \pm 1.2$   & $8.70  \pm 0.15$    \\
    \bottomrule
  \end{tabular}
  \caption{Fit parameters for the two exponential models.}
\end{table}

Both fits confirm that attractor length grows exponentially with $R$, in agreement with the theoretical prediction at the end of Section~\ref{attractorsec}. The model $\ell_{dom}(r)=r^a e^{b r}$ appears to match our selected points more closely, though this is influenced by our particular point–selection strategy. A pure exponential fit (as predicted in \cite{presuti1d}) could likely be obtained by including additional data, especially in the $m_0=0$ regime. 

\subsubsection{The role of system size on chimeras evolution. Metastable states}\label{sectionsize}

\begin{figure}[ht!]
\begin{center}
\includegraphics[width=\linewidth]{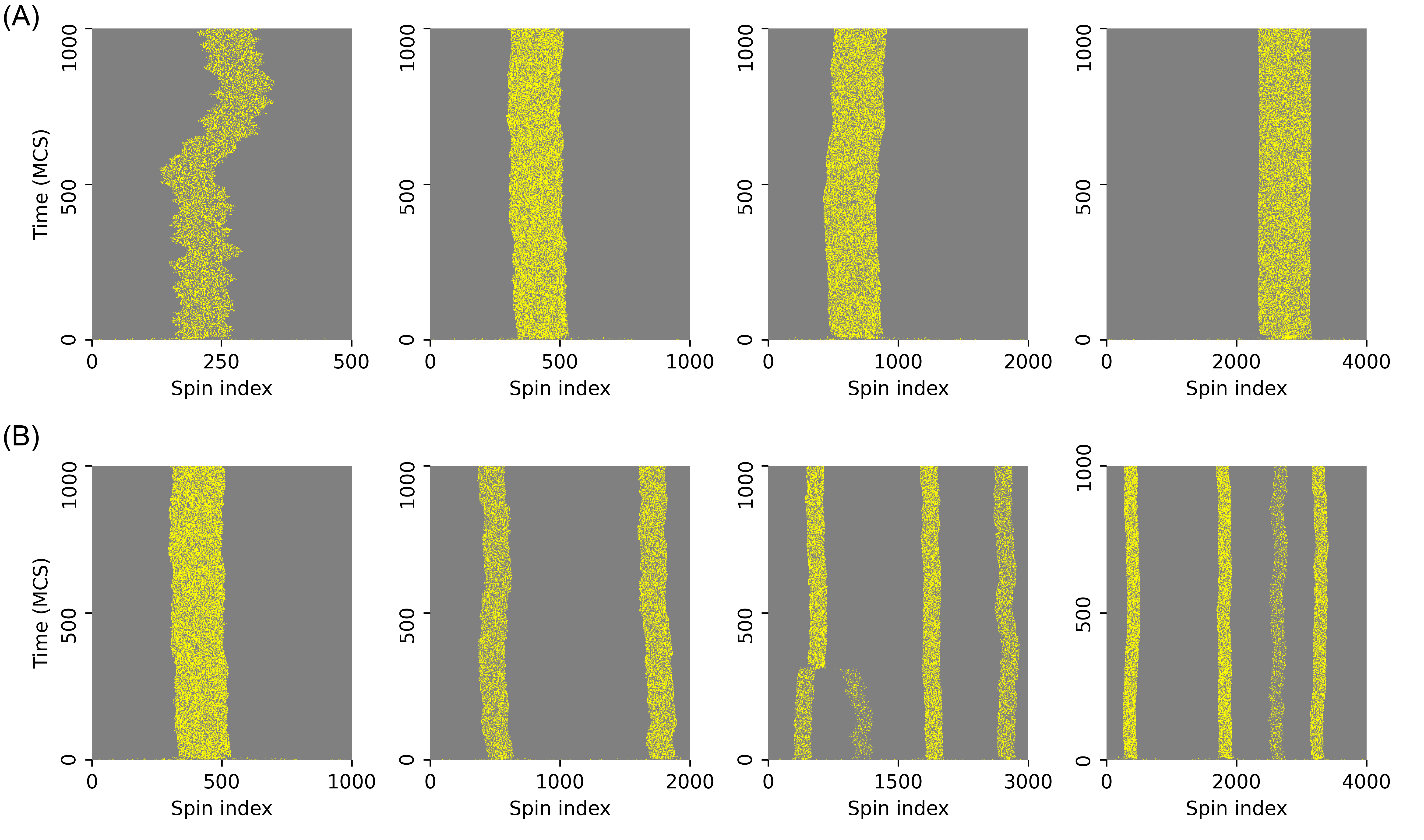}
\end{center}
\caption{\textbf{Dependence of chimera features on system size $N$ for $m_0\approx-0.8.$} Top panels (A): For a given $r=R/N=0.2$ the figure shows that chimera state remains as a stable attractor when $N$ increases so under this condition ($r=0.2$ fixed) one expects the chimera to remain in the thermodynamic limit. From left to right $N=500, \,1000,\, 2000$ and $4000$ respectively. It is also remarkable that the fluctuations of the borders decrease as $N$ increases. Bottom panels (B): These figures show the appearance of multichimeras for fixed $R=200$ and increasing values of $N.$ Note that the number of chimeras increases with $N$ but their width decreases with $N.$ Then, one would expect to have infinite number of chimeras with negligible width in the thermodynamic limit. From left to right $N=1000, 2000, 3000$ and $4000.$}
\label{sizecomparison}
\end{figure}

An interesting question to address is if the behavior observed in the systems for $N=512$ spins depends on the ring size, and what is more important, if such behavior remains when $N$ increases and still is present in the thermodynamic limit. To shed light about these questions we have performed a system-size analysis in Fig. \ref{sizecomparison}. Top panels of the figure show how a chimera changes when $N$ increases while keeping $r=R/N$ fixed. This analysis shows that chimera states do not disappear when $N$ increases and they will likely appear in the thermodynamic limit for $r$ fixed. Consequently chimera order can occur in an one-dimensional Ising model for $R\sim N.$ On the other hand, bottom panels of Fig. \ref{sizecomparison} show the behavior of chimera patterns when $R$ is fixed and $N$ is increased. The analysis shows that when $N$ increases the number of coexisting chimeras increases but the length of the chimera domains, relatively to the system size, decreases. This implies that short-range diffusion increases the probability of transient multichimera states. Moreover, this preliminary analysis indicates that it would be  expected to have an infinite number of very narrow chimeras in the thermodynamic limit. To validate this conclusion, we must first establish the scaling behavior of the number of chimeras with respect to $N$ in a numerical fashion.

Preliminary results in Fig. \ref{sizecomparison} show that the number of chimeras increase with $N$ and that through collisions they can form other chimeras or attractors. This last case occurs due to the fulfillment of condition (\ref{crit}) when $n^+$ is increased. So an interesting question to address is how the number of initials chimeras evolve in time until only attractors remain.
We have seen that attractors start to emerge when $n^+\approx R+1.$ Note moreover that $\langle n^+\rangle\approx (1+m_0)N/2\equiv q_0N$, Then, the condition for an attractor to emerge approximately is $q_0\approx(R+1)/N.$ 

Suppose that at any time $t$ we have a number of chimeras $N_c(t)$ with a {\em positive spins density} $\gamma(t)\le 1$ -- which is a measure of the fraction of positive spins within the chimera domain -- such that for $\gamma(t)=1$ one has an attractor instead of a chimera. Let also consider that initially a number $N_0=N_c(0)$ of chimeras emerges for a given $N.$  Each one of these chimeras typically will have a number of positive spins $n^+_c=\gamma_0{(R+1)}$ (with $\gamma_0=\gamma(0)< 1$) uniformly distributed over the range $R$, with  the rest negative spins. As a first approximation, we considered that all emerging chimeras have the same {chimera positive spin density} $\gamma_0$ and therefore the same $n^+_c.$ Then it follows that 
\begin{equation}
n^+=N_0 n^+_c =N_0\gamma_0(R+1)\approx q_0N.
\end{equation}
Consequently
\begin{equation}
N_0\approx\frac{q_0}{\gamma_0(R+1)} N\equiv\frac{\bar{\gamma}}{\gamma_0}N, 
\label{teornc}
\end{equation}
which indicates that  in the thermodynamic limit the initial number of chimeras scales with $N$ for $R$ finite ($\bar{\gamma}$ finite). Since we assumed that initially within a chimera one has $\gamma_0 (R+1)$ positive spins and $(1-\gamma_0)(R+1)$ negative spins, the mean length of these initial chimeras is 
\begin{equation}
L_{chimera}\approx R+1 
\end{equation}
independently of $\gamma_0.$ This is consistent with definition in Eq. (\ref{lengthchimera}) where now and in general for any time one has $|B_{n^+_c}(i_1,\ldots,i_{n^+_c})|=(1-\gamma(t))(R+1) $ and $n^+_c=\gamma(t)(R+1)$
We can conclude that initially 
\begin{equation}
N_c(0)=N_0\approx\frac{q_0}{\gamma_0 L_{chimera}} N. 
\end{equation}
It is worth noting that $\gamma(t)$, as said, is not constant and depends on chimera {\em observation} time $t_{obs}$ since we are considering metastable states. Due to the random walk nature of chimeras, they will eventually collide and give rise to a more dense chimera or to an attractor. Taking into account this consideration, the previous expressions are only valid for initial conditions of for a short initial amount of time.

\begin{figure}[ht!]
\begin{center}
\includegraphics[width=\linewidth]{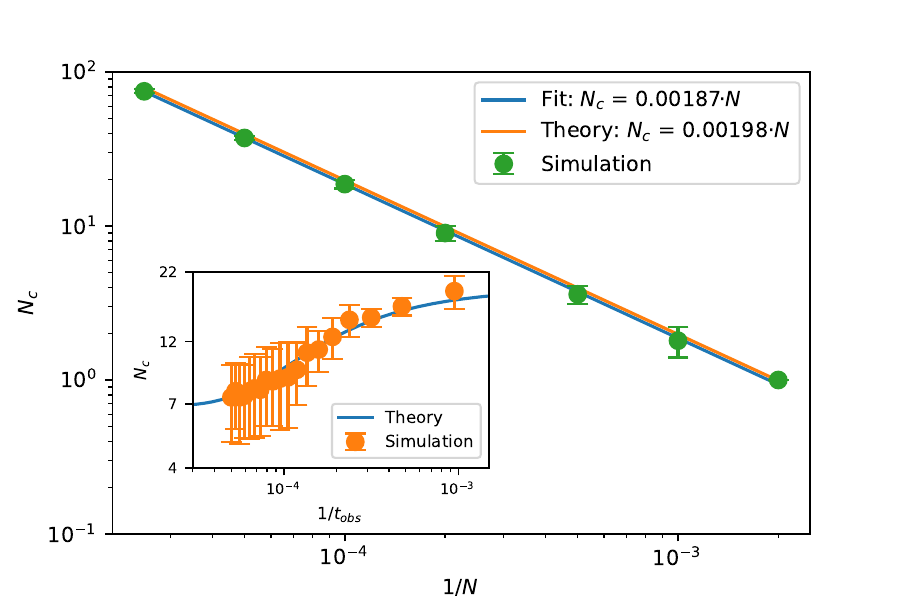}
\end{center}
\caption{Main plot: Dependence of the mean number of chimeras $N_c$ surviving at observation time $t_{obs}$ as a function of the system size $N$. Each data point has been obtained after averaging the measured $N_c$ over $10$ different simulations. Error bars have been computed using the standard deviation of these data. The figure shows a clear scaling as $N_c\approx \frac{q_0}{\gamma(t_{obs})(R+1)} N$ confirming the theoretical prediction in Eq. (\ref{teornc}) assuming $\gamma(t_{obs})=0.5$. Parameters  $t_{obs}=1000$ ~MCS, $q_0=0.1,$ $R=100$ and system sizes $N=500, 1000, 2000, 5000, 10000, 20000$ and $40000$ spins. Inset: Slow decrease of $N_c$  when the observation time $t_{obs}$ increases from $1$ to $2\times 10^4$ MCS for a system size of $N=10^4$. We observe a good agreement between simulation data and theoretical prediction in Eqs. (\ref{solfinal}) for $\tau_c=100$ and $\tau_a=200$ MCS. Each data point has been obtained after averaging over $10$  simulations.}
\label{numchimeras}
\end{figure}

Under these assumptions, we have tested the Eq. (\ref{teornc}) numerically in Fig. \ref{numchimeras} (Main plot) for a relatively short observational time of $t_{obs}=1000$ MCS.  The figure shows a clear scaling of the number of surviving chimeras at this time with the system size, while its inset indicates that when the observation time increases the number of existing chimeras decreases. 
Note, moreover, that the initial number of chimeras in the thermodynamic limit can already be established by Eq. (\ref{teornc}) just taking $R=rN$ which yields $N_0\approx q_0/\gamma_0 r$, expression  that only depends on intensive quantities. 

When considering large observation times, one has to consider that pairs of neighboring chimeras can collide and originate attractor or denser chimeras, which have the same length $R+1$. In this situation, one can define the following dynamics for the evolution of the number of chimeras in time:
\begin{equation}
\tau_c\frac{dN_c(t)}{dt}=-2p_0f(\gamma(t))N_c (t) n_{vec}\frac{R+1}{N}
\label{teornctime}
\end{equation}
{with $\tau_c$ the typical time scale for evolution of chimeras. This equation gives the decrease of the number of chimeras in time (see the negative sign in the right-hand side of the equation). Since a decrease in the number of chimeras mainly occurs because collisions between chimeras generate denser chimeras or attractors, it will depend first in how fast chimeras move in both directions of the ring (this is why the factor 2 appears after the minus) which is related with the probability $p_0=1/2$ of moving in each direction of the ring. This probability is multiplied by the factor $f(\gamma(t))$ which  accounts for a normalized factor that decreases the probability  of displacement $p_0$ as if it was a random walker in an increasingly crowded environment of positive spins (see Section \ref{mdlsst}), and therefore decreasing the probability of one chimera to collide with other chimeras.
It is sufficient for the purpose of present analysis to assume that it is a decreasing function of $\gamma(t)$ for $\gamma(t)\in (0,1)$ and satisfies $f(1)=0$, as that would be the attractor scenario. Finally, the decrease of the number of chimeras also will depend on the number of pairs of neighboring chimeras at given time, i.e. the factor $N_c(t) n_{vec}$ in the right-hand side of equation (\ref{teornctime}), with $n_{vec}$ being the number of neighboring chimeras to a given one (with possible values $n_{nec}=0,1,2$) and on the fraction of the ring that a chimera occupies, i.e. the last factor $\frac{R+1}{N}$ on the right-hand side of Eq. (\ref{teornctime}).}

Additionally, the number of attractors $N_a(t)$ also will evolve in time from an initial value $N_a(0)=0$. Attractors emerge when two very dense chimeras collide, so the attractor dynamics depends on $N_c(t)$. Then one can assume the following equation for the evolution of attractors:
\begin{equation}
\tau_a\frac{dN_a(t)}{dt}=p_0f(\gamma(t))N_c n_{vec}\frac{R+1}{N}
\label{teornatime}
\end{equation}
{with $\tau_a$ the time constant for the evolution of attractors. The reason for the choice in the right-hand of Eq. (\ref{teornatime}) is that the number of atractors increase (possitive right hand side of the equation) in the same way that chimeras decreases but its number increases at a rate that is just half of the decrease of chimeras since an attractor generally originates when two chimeras collide. One has to define also a time scale of the evolution of attractor $\tau_a$ different than the time scale for the evolution of the chimeras $\tau_c$ since not all chimera collisions result in an attractor so $\tau_a>\tau_c$. We can see that $N_a(t)$ will affect  indirectly the dynamics of chimeras through $\gamma(t)$ if we assume that for any time $N_c(0)\gamma_0=N_a(t)+N_c(t)\gamma(t)=\bar{\gamma}N$ (see above). }

Assuming the choice $f(\gamma(t))=1-\gamma(t),$ Eqs. (\ref{teornctime}) and (\ref{teornatime}) become
\begin{equation}
\begin{array}{c}
\displaystyle\tau_c\frac{dN_c(t)}{dt}=-2p_0n_{vec}[N_c(t) + N_a(t) -\bar{\gamma}N]\frac{R+1}{N}\\
\\
\displaystyle\tau_a\frac{dN_a(t)}{dt}=p_0n_{vec}[N_c(t) + N_a(t) -\bar{\gamma}N]\frac{R+1}{N}.\\
\end{array}
\label{teornctime2}
\end{equation}

Considering now the initial conditions $N_a(t=0)=0$ and $N_c(t=0)=N_0$ one finally obtain the solution
\begin{equation}
\begin{array}{ll}
\displaystyle N_c(t) = \frac{2\tau_a S(t) - \tau_c N_0}{2\tau_a - \tau_c}
\\
\\
\displaystyle N_a(t) = \frac{\tau_c \left[ N_0 - S(t) \right]}{2\tau_a - \tau_c}
\end{array}
\label{solfinal}
\end{equation}
with
\[
S(t) \equiv N_c(t)+N_a(t)= \bar{\gamma} N + (N_0 - \bar{\gamma} N) e^{\alpha t},
\]
where
\[
\alpha = K\left( \frac{1}{2\tau_a} - \frac{1}{\tau_c} \right), \quad 
K = \frac{2p_0 n_{\text{vec}}(R+1)}{N}.
\]
Only when $\alpha<0$ a real solution exists which implies $\tau_c<2\tau_a.$ Note that with this solution one can have $N_c(\infty)=0$ only if $2\tau_a \bar{\gamma}N=\tau_c N_0$, for which the solution becomes
\begin{equation}
\begin{array}{lll}
\displaystyle N_c(t) &=& \displaystyle N_0 \, e^{\alpha t} \\
\\
\displaystyle N_a(t) &=& \displaystyle\frac{\tau_c N_0}{2\tau_a} \left( 1 - e^{\alpha t} \right).
\end{array}
\end{equation}

We have tested the validity of this last expression in the inset of Fig. \ref{numchimeras}, finding a very good agreement of the theoretical line with simulations for $\tau_c=100$ and $\tau_a=200$ MCS, which in this way gives a typical and average time scale for merging chimeras and appearance of attractors, respectively. These values are reasonable since the appearance of attractors will take longer times. 

\subsubsection{Phase diagram of the system}\label{mdlsst}
\begin{figure}[ht!]
\begin{center}
\includegraphics[width=\linewidth]{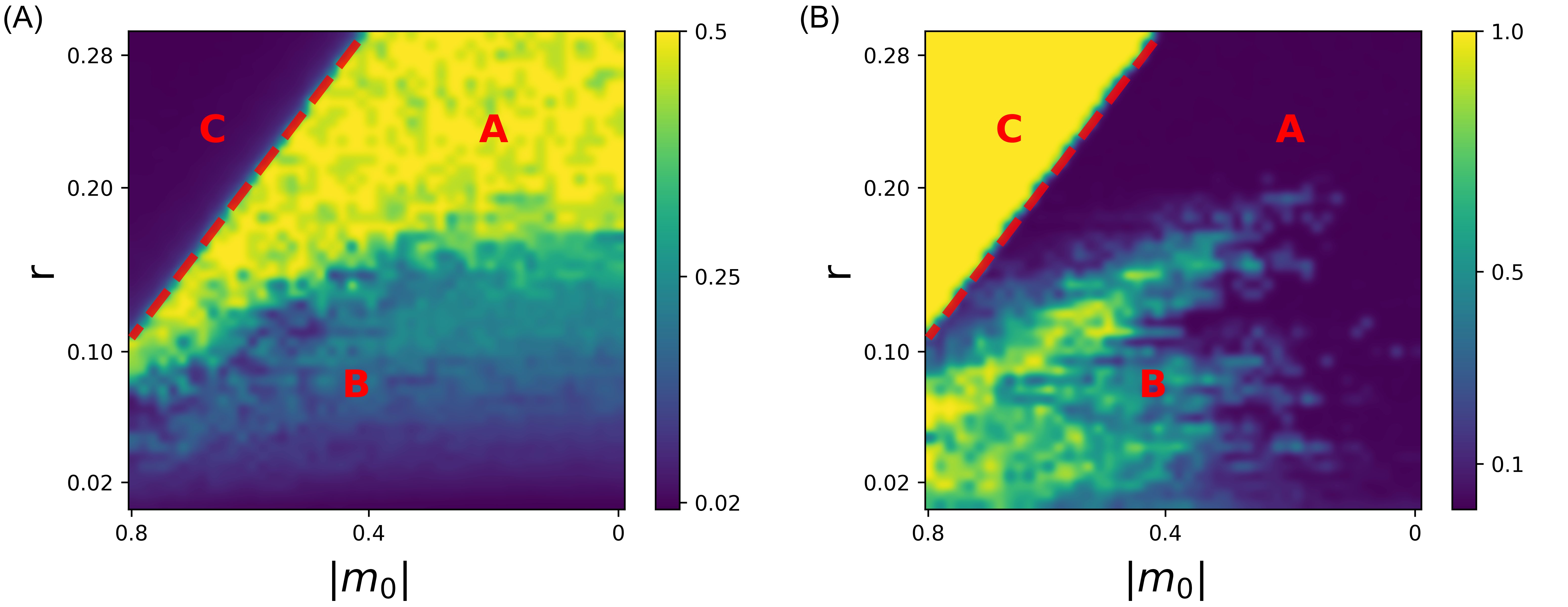}
\end{center}
\caption{\textbf {Phase diagram of the system for a ring of $N=512$ spins.}
(A) Normalized Mean Domain Length $\ell_{dom}$ and  (B) Normalized Stationary State Time $\tau_{ss}$ both as a function of $r$ and $|m_0|$. Region A: pure attractor states. Region B: coexisting chimeras and attractors. Region C: pure chimera states.}
\label{figure5}
\end{figure}

We have investigate here how the emergent behavior of our system changes with relevant parameters. To map the full parameter space of our system, we computed $\ell_{dom}$ and $\tau_{ss}$ for a wide range of $r$ and $|m_0|$ pairs, averaging over 10 simulations.  Fig. \ref{figure5}(A) shows the resulting phase diagram for \emph{Normalized Mean Domain Length}, and Fig. \ref{figure5}(B) shows the \emph{Normalized Stationary State Time} $\tau_{ss}$ phase diagram, where we have chosen to represent $|m_0|$ to stress that the results are symmetric for negative and positive values of the magnetization $m_0$. 

The analysis of both phase diagrams shows the emergence of three main regions: Region A which corresponds to $\ell_{dom}$ $\approx 0.5$, i.e., wide attractors, with very short $\tau_{ss}$, corresponding to two alternating domains (positive and negative) that form rapidly. Secondly, Region B where $\ell_{dom}$ declines gradually and $\tau_{ss}$ increases gradually with $r$, indicating the presence metastable chimera domains coexisting with attractors. Finally, Region C is characterized by an abrupt drop in $\ell_{dom}$ and abrupt increase in $\tau_{ss}$, matching the $|m_0| = 2r-1$ criterion from Eq. (\ref{crit}), as represented by the dashed red line. In Region C, attractors can not exist so only chimeras remains. These numerical results confirm the analytical predictions (see also Eq. (\ref{crit})) for attractor formation, as well showing the metastable region discussed in Section \ref{sectionsize}.

\paragraph{Random‐walk picture of metastable chimeras.}
The long‐lived \emph{metastable} chimera states in Region~B can be understood by mapping the noisy domain of \(n^+\) positive spins onto a constrained random walk of \(n^+\) walkers within the overlapping ``ball'' \(B_{n^+}\) (see definition (\ref{balln})).  At zero‐temperature, each positive spin executes exchanges with an available negative neighbor in \(B_{n^+}\) with equal probability (see Eq. (\ref{eq:Kawasakin})).  
As \(n^+/R\) increases ($|m_0|$ decreases) for a fixed $r$, $ |B_{n^+}| $ shrinks in size relative to \(n^+\) and one arrives to a more ``crowded'' random‐walk habitat: walkers (positive spins) must coordinate over fewer available sites, and their joint diffusion slows down. Chimeras would then take longer to find another chimera (forming a domain that, in many cases, is larger than $R+1$ and then would decay into an attractor instead) or find an attractor and being absorbed into it.  The behavior in Region B can be described in terms of a \emph{mesoscopic scale}: the system would first organize into moving chimeras and fixed attractors (depending on the number of positive spins that are in a certain medium size or mesoscopic region) and then these chimeras would move, interacting with other moving chimeras or some attractors.

\subsubsection{Phase diagram: Additional measures to characterize chimera‐state observables}
\begin{figure}[ht!]
 \begin{center}
     \includegraphics[width=\linewidth]{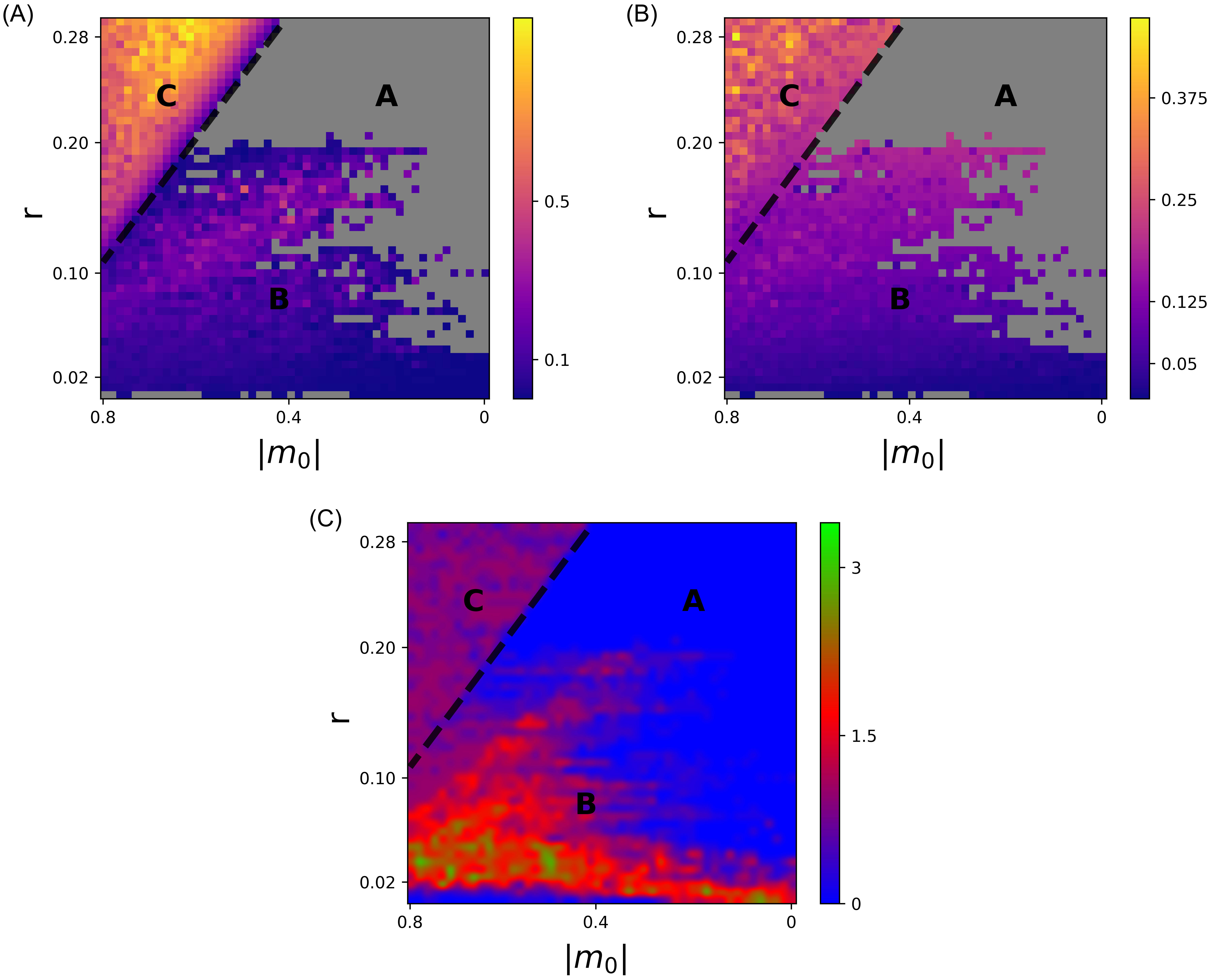}
 \end{center}
  \caption{Chimera‐state metrics across the \((r,|m_0|)\) phase diagram. (A) Normalized Mean Activity, (B) Mean Peak Width divided by $N$ and (C) Mean Number of Chimeras. Gray regions denote parameter combinations in which no sustained chimera events were detected.}
  \label{figure6}
\end{figure}

In region B of the \((r,|m_0|)\) plane, attractors and chimera domains coexist in a richly fluctuating, metastable regime.  Although we have seen that $\tau_{ss}$ can detect if an attractor is present o not in the system, to better quantify the chimera component in such region, we analyzed in Fig. \ref{figure6} three complementary observables from each simulation. First, in  Fig. \ref{figure6} (A) we computed the {\em Normalized Mean Activity}. Initially, we measure the \emph{activity} at site \(i\) as in Fig. \ref{types}. Peaks in this activity profile identify chimeras. Averaging the height of these peaks across all chimeras (and over $10$ simulations) yields the mean activity, which was further normalized to the maximum height of a single chimera. We see that in region C, a single, large chimera appears in the ring, giving rise to a consistently high activity peak. In Region B, multiple smaller chimeras form and dissolve, so the mean activity is lower and more variable. Region A corresponds to pure attractors with not chimera activity (lack of color data points). Secondly in Fig. \ref{figure6} (B) we measured the \textit{Mean peak width} of chimeras states. For each detected chimera activity peak we measure its full‐width at half‐maximum (\textit{FWHM}), normalized by \(N\). Then we average these quantities  over all detected chimeras and $10$ simulations.  The resulting mean width reflects how spatially extended each typical chimera is on a particular phase diagram point. The resulting plot does not show a clear distinction between regions B and C, but a continuous increase in chimera length. This is consistent with the increase of chimera length with $R$, which expands the number of sites with which positive spins can exchange in a linear manner, as we discussed in the previous section. Finally, in Fig. \ref{figure6}(C), we analyzed the \textit{Mean number of chimeras} that emerge for each point of the ($r,|m_0|$) plane. We count the number of distinct activity‐peak events per simulation,   which is equal to the number of coexisting chimeras.  Averaging across simulations, we find that Region C is characterized by the emergence of  one dominant chimera (value $\approx$ 1). Region B typically sustains \emph{two to three} chimeras simultaneously, reflecting an intermediate mesoscopic scale in which chimeras can exist (and co-exist with attractors), and finally Region A clearly does not shows chimeras as expected.  

In summary, together, these three observables paint a clear picture of how chimera emergence occurs across the plane ($r,|m_0|$). In {Region C}, the system settles into one broad, highly active stable chimera that persists indefinitely. 
In Region B, multiple chimeras form.  The fact that they mostly coexist in an attractor landscape makes them less stable in time (as shown for example in Fig.~\ref{types})(D).

\subsubsection{Characterization of attractors across the phase diagram}
\begin{figure}[ht!]
\begin{center}
    \includegraphics[width=\linewidth]{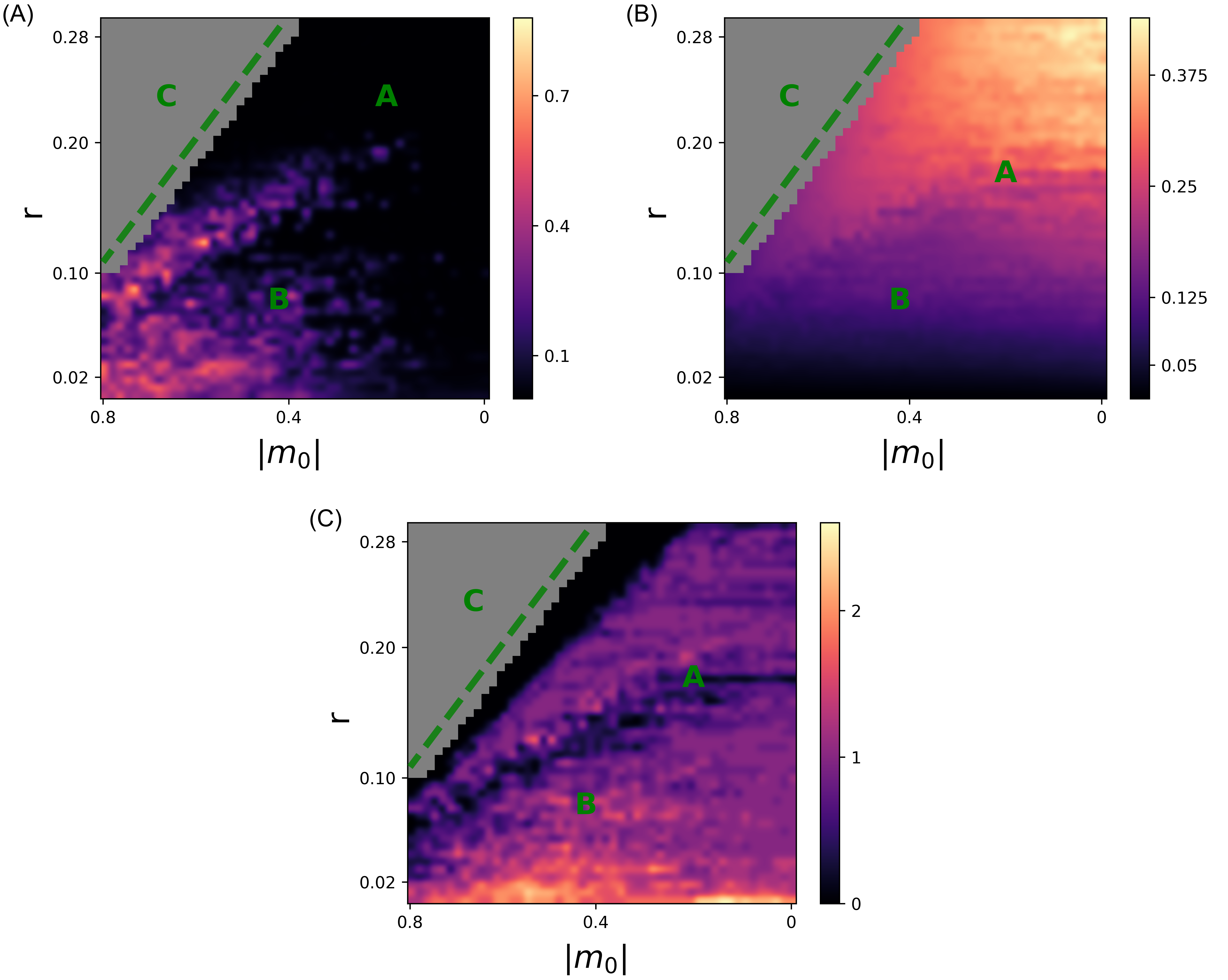}
\end{center}  \caption{Attractor state metrics across the \((r,|m_0|)\) phase diagram. (A) Normalized Maximum Attractor formation Time. (B) Mean Attractor Width. (C) Sudden Attractor Coarsening Events.}
  \label{figura7}
\end{figure}

We have also analyze the main characteristics of  attractors in the system when one varies $r$ and $m_0.$ The results of this analysis are shown in Fig. \ref{figura7}.

Across the \((r,|m_0|)\) plane, we monitor attractor formation and evolution through three key observables. First, the \textit{Maximum Attractor formation Time} illustrated in Fig.~\ref{figura7} (A). For each simulation we recorded the largest normalized MCS at which an attractor appears.  In Region A, these times are close to $0$, which reflects their rapid formation. In Region B, formation times increase with $|m_0|$ and decrease with $r$. This suggests that attractors can appear in this region as the consequence of the interaction of chimeras, as shown in  in Fig.~\ref{types})(D). Secondly, we computed the \textit{Mean Attractor Width} in Fig.~\ref{figura7} (B), by measuring the spatial extent of the largest attractor normalized by \(N\) and averaging over $10$ simulations. In this case we see no distinction between Regions A and B, reflecting a continuous increase with $r$, as discussed in detail in Section \ref{length}, and a continuous decrease with $|m_0|$, as the number of positive spins grows. Finally in Fig. \ref{figura7} (C) we measured the \textit{Sudden Attractor Coarsening Events} by counting the number of times a given attractor increases in size by a factor of at least $1.2$ over less than $100$ MCS. We see once again a continuous change from Regions A to B, with the probability of a coarsening event having a stronger dependence with $r$ than with $|m_0|$, probably accounting for the fact that there are simply more attractors the smaller $r$ is.

After the careful analysis performed in Fig. \ref{figura7} we can conclude  that in {Region A}, attractors form rapidly and remain stable, with continuous width dependence of the parameters and early coarsening events. Moreover, in {Region B}, attractor formation is in some cases delayed because of interactions of previous chimeras, widths are smaller, and sudden coarsening events due to chimeras merging into them are more frequent.

This thorough analysis of computational results of our system establishes a bridge between the rigorous analytical predictions and the  behaviors revealed by simulations.

\section{Conclusions}
In this work, we have demonstrated via Monte Carlo simulations and using analytical derivations that a one-dimensional Ising model with long-range interactions (i.e. $r=R/N$ finite in the thermodynamic limit) supports chimera states under purely diffusive dynamics in a hamiltonian system that obeys detailed balance. To our knowledge, this represents the first description of chimeric behavior in the classical definition of an Ising system in which all spins are identical and equally coupled to each other. This demonstrates that,  for such intriguing behavior to emerge, it is not necessary to include distinct couplings in subdomains of the system, as it was reported in a previous work \cite{Singh2011}. 

We have further demonstrated, both analytically and through simulations, that when $R$ is fixed and $N$ increases, chimera-like states still emerge, and some of them persist throughout the maximum simulation time recorded. In this case, as $N$ increases, the initial number of chimeras also increases, but their lengths decrease with $N$. During the simulations, the number of chimeras gradually decreases due to a merging processes among them, leading to the emergence of attractor  states. We have also theoretically demonstrated this behavior using simple dynamical assumptions that describe the chimera collision process.

Our analysis reproduces certain results previously reported in the literature concerning the emergence of attractors -- e.g., in \cite{presutti, presuti1d} -- as spin interactions represent a specific instance of Kac potentials in the one-dimensional Ising model, which we have derived in Section \ref{attractorsec}.

For the case of emergent stable chimeras (finite $r$ in the thermodynamic limit), we explored the full range of relevant system parameters and identified three distinct phases, as illustrated in the 
($r, |m_0|$) plane. Multiple metrics were tested, all providing a consistent characterization of these regimes, as detailed in Section \ref{mdlsst} and beyond.
Moreover, we derived an empirical relation that defines the conditions under which only chimera states can exist, providing a clear boundary for their appearance in the ($r,|m_0|$) phase plane. This boundary closely matches the results from our simulations, as reported in Section \ref{analyt}.

Interestingly, numerical simulations revealed a region of the phase space where chimeras and attractors coexist and interact, often leading to the eventual collapse of chimeras into stable attractors. This transition can be clearly observed by fixing $R$ and increasing $N$ which effectively reduces $r.$ In doing so, the system passes from the region of stable chimeras to that of chimera-attractor coexistence.

An essential factor in the emergence of chimera states is that the Metropolis transition rate (Eq. (\ref{eq:metropolis})) inherently accepts zero-energy moves. In classical systems, however, any transition involves a finite energy cost, raising the question of why such moves are accepted. The simulated dynamics should therefore not be interpreted as physically realizable. Rather, unconditional acceptance of transitions between configurations with identical energy (and thus identical Boltzmann weights, $e^{-\beta \mathcal{H}}$) ensures unbiased sampling of the configuration space. Importantly, the main result -- the existence of stable chimera states in the one-dimensional Ising model with long-range interactions -- remains robust when other transition rates, such as the Glauber dynamics, are employed (data not shown).

In conclusion, we have shown that a uniformly coupled, one-dimensional Ising system supports the existence of chimera states under diffusive dynamics in the thermodynamic limit. {Although we have performed our study in noiseless conditions ($T=0$) where simple analytical derivations can be obtained, preliminary simulations shows that most of the behaviour reported in the present work also holds for $T>0$, including a rich phase diagram with the appearance of stable chimeras and attractors but now with boundaries between chimeras and attractors more diffuse and with the convergence of chimeras into attractors occurring quickly. However, we think that the extension of the present work for $T>0$ is more appropriate for a future work since it requires an in-depth analysis and the development of different types of analytical approaches for the proper characterization of the system’s emergent behavior.}
Another interesting question to address is if such intriguing chimera patterns can also appear and be stable in non-equilibrium situations, for example including reaction and diffusion competing dynamics. Extending the present study for such situations could be of great interest. In fact, some preliminary simulations of such dynamics point out in this direction (data not shown). We think, however, that a fully study of this non-equilibrium system setup is beyond the scope of the present study and be more appropriate for a future work. 

Finally, given the extensive use of the Ising model for the study of many different complex systems, our findings in the present work offer a solid basis for exploring chimera theory in such systems. These results also point to potential applications in systems with discrete collective behavior, such as complex neural networks, therefore suggesting promising applications in neuroscience. 

\section{Acknowledgments}
This work has been supported by Grant No. PID2023-149174NB-I00 financed by the Spanish Ministry and Agencia Estatal de Investigación MICIU/AEI/10.13039/501100011033 and ERDF funds (European Union). We thank Christian Maes and Víctor Buendía for their valuable feedback on this research, which helped enhance both the clarity and depth of the present work. 

\bibliographystyle{unsrt}
\bibliography{references}

\end{document}